\documentclass[aps,prc,groupedaddress,showpacs,twocolumn,floatfix]{revtex4}

\usepackage{graphicx}
\usepackage[dvips]{color}
\usepackage{colordvi}

\def\gtorder{\mathrel{\raise.3ex\hbox{$>$}\mkern-14mu
 \lower0.6ex\hbox{$\sim$}}}
\def\ltorder{\mathrel{\raise.3ex\hbox{$<$}\mkern-14mu
 \lower0.6ex\hbox{$\sim$}}}

\def\beq{\begin{equation}}
\def\eeq{\end{equation}}
\def\ba{\begin{eqnarray*}}
\def\ea{\end{eqnarray*}}

\begin{document}

\title{Zemach moments of $^3$He and $^4$He}

\author{Ingo Sick }
\affiliation{Dept.~f\"{u}r Physik, Universit\"{a}t Basel,
CH4056 Basel, Switzerland}

\date{\today}
\vspace*{5mm}

\begin{abstract}
 We use the {\em world} data on elastic electron scattering on
$^3$He and $^4$He to determine the Zemach moments $\langle r \rangle _{(2)}$ and
 $\langle r^3 \rangle _{(2)}$. These quantities are required to
interpret the Lamb shift and HFS data of muonic Helium presently being
measured  at PSI by the CREMA collaboration. The {\em rms}-radii are determined as
well.
\end{abstract}


\email{ingo.sick@unibas.ch}
\maketitle

\noindent {\em Introduction. ~}
In this note, we present results on the rms-radii and Zemach moments 
of the Helium isotopes $^3$He and $^4$He. 
The interest in these integral quantities is threefold: 

1. Precise moments are useful observables for the comparison with theoretical
calculations. This is true in particular for light nuclei such as the Helium
isotopes where very accurate {\em ab-initio} calculations can be performed.

2. At present there are experiments underway to measure the charge 
rms-radii of the Helium
nuclei via the Lamb shift  in the muonic Helium ion \cite{Antognini11}. 
For the interpretation of
these data --- which will ultimately provide rms-radii that are much more
precise than the ones extracted from electron scattering ---  corrections
depending on the Zemach radii are needed \cite{Friar04,Friar05b}. 
These quantities can be determined via electron scattering.

3. There is presently a major discrepancy between the rms-radius of the proton
as determined from electron scattering \cite{Sick12} and muonic Hydrogen
\cite{Pohl10a}, respectively. One of
the speculations concerning the origin of this discrepancy involves a potential 
difference
in the ''electromagnetic'' interaction between electrons and muons. It  is then
desirable to make a comparison between radii from experiments involving 
$e$ and $\mu$ for other cases. The
most accurate confrontation can be performed for $^4$He, the nucleus for which
the relative uncertainty of the rms-radius from electron scattering is smallest.

We also note that the  measurements in  the (electronic) Helium atom of the
$^3$He--$^4$He isotopic shift differ by several standard deviations. It is of
interest to see whether electron scattering can help to resolve the
issue. \\[5mm]

\noindent {\em Moments for $^4$He. ~}
For the interpretation of the Lamb-shift data for  muonic $^4$He, which are
presently being taken by Antognini {\em et al.} at PSI, the third Zemach moment   
is needed in order to extract the rms-radius. 
 This moment can be computed \cite{Friar04,Friar05b} from the charge form factor $G_e(q)$ depending on
momentum transfer $q$ 
\ba \langle r^3 \rangle_{(2)} = \frac{48}{\pi} \int_0^\infty \frac{dq}{q^4}~
(G_e^2(q)-1+q^2 R^2/3)
\ea
where $R$ is the charge rms-radius. 

In \cite{Sick08} we have performed a Sum-Of-Gaussians (SOG) fit to the {\em
world} data on elastic electron scattering from $^4$He 
\cite{Frosch67}-\nocite{Erich68,McCarthy77,Arnold78,vonGunten82}\cite{Ottermann85}.
 We have, for completeness, 
added the recently published high-$q$ data of Camsonne {\em et al.} 
\cite{Camsonne13} and redone the fit. For details see \cite{Sick08}. 

The resulting charge rms-radius is (as in \cite{Sick08}) 1.681 $\pm$0.004$fm$.
The third Zemach moment is found to be 16.73$\pm$0.10$fm^3$, where the error
bar covers both the random and systematic uncertainties of the data.
 For comparison:  for Gaussian (exponential)  densities   --- which are often 
used to estimate $\langle r^3 \rangle _{(2)}$ --- this moment, for the same 
rms radius,  would amount to 16.50 (17.99) $fm^3$. 

One should note that the appearance of the $1/q^4$ factor in the expression for 
$\langle r^3 \rangle_{(2)}$ does not imply that this moment depends strongly on
the (e,e) data at extremely low $q$. The low-$q$ dependence of 
$G(q) \sim 1-q^2 R^2/6+...$ cancels the "$-1+q^2R^2/3$" term.
In fig. 1 we show the convergence of the Zemach integral as a function of the
upper integration limit. While the full curve
gives the Zemach integral (which converges very slowly),
 the dashed curve has the integral  over
the formfactor-independent term $(-1+q^2R^2/3)/q^4$ up to $q=\infty$ added in. 
These curves  show that the experimental information on $G(q)$ 
in the entire region $0 \div$1 fm$^{-1}$ contributes; above $q \sim 1.2
fm^{-1}$ $G(q)$ is too small to contribute substantially.

\begin{figure}[bht]
\begin{center}
\includegraphics[scale=0.48,clip]{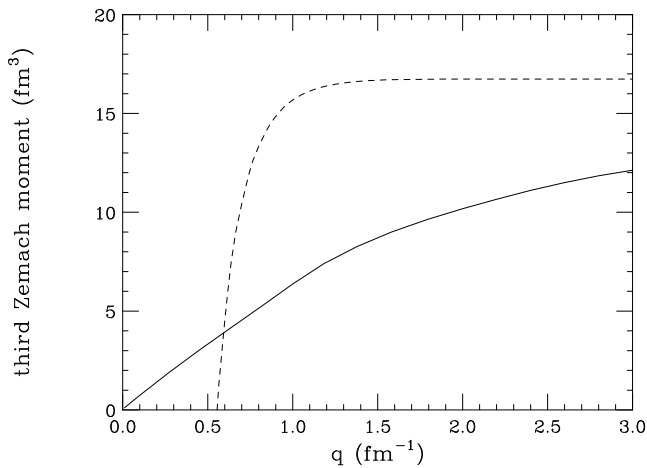}
\parbox{8cm}{\caption[]{Convergence of the integral for $\langle r^3
\rangle_{(2)}$ }} 
\end{center} 
\end{figure}
 
The $q$-region of sensitivity to $\langle r^3 \rangle_{(2)}$ turns out to be
quite similar to the one for the rms-radius and the first Zemach moment 
to be discussed below. 

For some applications it might also be useful to have the fourth moment
$\langle r^4 \rangle$. It amounts to 14.35$\pm$0.11$fm^4$. The various moments
are summarized in table 1. 

\begin{table}[htb]   
\hspace*{0.5cm} \begin{tabular}{l|r}
$\langle r^3 \rangle_{(2)}$ &  ~~$16.73 \pm 0.10 fm^3$ \\
\hline $\langle r^2 \rangle^{1/2}$ &  $1.681 \pm 0.004 fm$ \\
\hline $\langle r^4 \rangle $ & $14.35 \pm 0.11 fm^4$ \\
\end{tabular} 
\parbox{6cm}{ \caption{Moments for $^4$He}}
\end{table}   

\noindent {\em Moments for $^3$He.  ~}
As Antognini {\em et al.}  are studying muonic $^3$He as well, we have performed a
similar analysis of the {\em world} data for $^3$He. For this nucleus a less
extensive set of data is available 
\cite{vonGunten82,Ottermann85}, \cite{Szalata77}%
\nocite{Dunn83,McCarthy70,McCarthy77,Cavedon82,Nakagawa01a,Arnold78,Collard65}%
-\cite{Beck87a}.  The data are in general not as precise as for $^4$He. A
 complication arises from the spin-1/2 nature of $^3$He. In this
case the data depend on {\em two} quantities, the charge (monopole) and the 
magnetic  (dipole) form
factors $G_e$ and $G_m$, respectively. As both forward-  and backward-angle 
data are available, these
form factors can  be separated, at the expense of an increase of the 
uncertainties.   Figure 2 shows the 
low-$q$ data which are of special interest for the determination of the moments.

\begin{figure}[bht]
\begin{center}
\includegraphics[scale=0.48,clip]{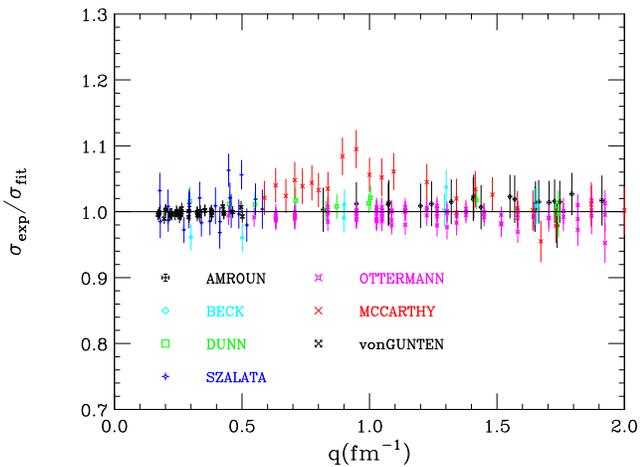}
\parbox{8cm}{\caption[]{(Color online) Ratio of experiment to fit for $^3$He. Data with 
$\delta \sigma / \sigma > 0.1$ are not shown.}} 
\end{center} 
\end{figure}

In order to determine the form factors, we have fitted the {\em world} data set,
mainly in the form of unseparated cross sections, using the SOG
 parameterization for $G_e(q)$ and $G_m(q)$, hereby yielding the optimal e/m separation.
 The data
were corrected for Coulomb distortion if this had not been already done by the
authors. Random errors of the derived quantities were determined using the error
matrix, the systematic errors of the data (mainly normalization) were included
by changing the data sets by the quoted error, refitting and adding
quadratically all the resulting changes. The SOG fit of the {\em world} data, comprising 354 data points 
up to $q_{max}=10 fm^{-1}$, has a $\chi ^2$ of 346. The results for the third 
Zemach moment and the rms-radii are listed in table 2.

As for $^4$He \cite{Sick08} the large-radius tail of the density has been
constrained to have the fall-off as given by the proton separation energy (modulo
corrections which are of minor quantitative impact). For the determination of 
the rms-radius the knowledge on the {large-$r$} behavior of
$\rho (r)$ is important to bridge the gap between the region of
 $0.5<q<1.2 fm^{-1}$ where the
data are sensitive to the rms-radius to the $q=0$ point 
where the rms-radius is obtained from 
the slope of $G_e(q)$ as function of $q^2$ \cite{Sick14}. Also for the Zemach
moments the information on the large-$r$ shape of $\rho (r)$ removes
the major source of model dependence inherent in
the choice of the parameterization for $\rho (r)$ or $G_e (q)$. 

Contrary to the case of $^4$He we
do not know the absolute value of the density \cite{Sick08}, so only the 
{\em shape} of $\rho (r)$ can be added as additional input.  This shape  
 is fitted for radii where the nucleon wave functions
are in the asymptotic regime, {\em i.e.} where they are outside the nuclear
potential and fall like a Whittaker function depending on the nucleon separation
energy.  This is the case for radii where the density typically has fallen to
less than 1\% of the density in the nuclear interior. 

In order to explore a potential model dependence introduced by this procedure, we
have compared the radii determined using shapes from rather different sources.
On the one hand side, we have used the shapes from densities from
GFMC \cite{Brida11} and Faddeev \cite{Sauer00} calculations performed using modern 2N- and
3N-potentials. As an alternative, we have calculated the p and n densities in a
Woods-Saxon potential fitted to the form factor. The corresponding
point densities have been folded with the nucleon  densities and added.
Also in this case, the \mbox{large-$r$} behavior is given entirely by the p- and
n-separation energies, which are accurately known from experiment.
The comparison of the resulting moments shows no significant dependence on the
tail-shape used.

For a spin-1/2 nucleus such as $^3$He it is also of interest to compute the 
standard (first) Zemach moment 
which can be obtained from the form factors via 
\ba
\langle r \rangle_{(2)} = - \frac{4}{\pi} \int_0^\infty (G_e(q)~ G_m(q)-1)
\frac{dq}{q^2} ,
\ea
where $G_m(q)$ is the magnetic form factor (normalized at $q=0$ to 1).
This moment is needed to compute the finite size effects in the 
hyperfine splitting in muonic atoms, a quantity also being measured by the CREMA
collaboration. One could naively have  expected that the
HFS would basically depend on the  magnetization density  $\rho_m(r)$ alone. 
The actual situation is somewhat more complicated as the lepton wave function 
inside the
nucleus is influenced by the distribution of the charge. 

\begin{table}[htb]
\begin{tabular}{l|r}
$ \langle r \rangle_{(2)}$ & ~~~$2.528 \pm 0.016 fm$ \\
$ \langle r^3 \rangle_{(2)}$ & $28.15 \pm 0.70 fm^3$ \\
\hline
$ \langle r^2_{ch} \rangle ^{1/2}$ & $1.973 \pm 0.014 fm$ \\
$ \langle r^2_{m} \rangle ^{1/2}$ & $1.976 \pm 0.047 fm$ \\
\hline
$ \langle r^4_{ch} \rangle $ & $32.9 \pm 1.60 fm^4$ \\ 
\end{tabular}
\parbox{6cm}{\caption{Moments for $^3$He}}
\end{table}

The results for $^3$He:  $\langle r \rangle_{(2)} = 2.528 \pm 0.016 fm$, $
\langle r^3 \rangle_{(2)} = 28.15 \pm 0.70 fm^3$. For  Gaussian (exponential)
densities --- which are often used to estimate the Zemach moments ---    
$ \langle r^3 \rangle_{(2)}$   with the rms-radius R of the SOG fit would 
amount to 26.68(29.10)$fm^3$;  $\langle r \rangle_{(2)}$, with both radii set to the
experimental charge radius,  would amount to  2.570(2.492)$fm$.   \\[5mm]

\noindent {\em Isotope shift. ~}
 From the charge radii of $^3$He and $^4$He listed above we deduce  
 an isotope shift
$^3$He--$^4$He of $\delta \langle r^2 \rangle$ = 1.066$\pm$0.06 $fm^2$.
This shift can be compared to values  
\cite{VanRooij11}-\nocite{Pastor12,Pachucki12}\cite{Shiner95}
determined in atomic (electronic) Helium.

Shiner {\em et al.} measured the $2^3 S_1$--$2^3 P_0$ transition in $^3$He, 
Van Rooij {\em et al.} observed the orthohelium-parahelium, doubly forbidden 
 transition between
the metastable $2^3S_1$ and $2^1S_0$ states in $^3$He and $^4$He. 
Cancio Pastor {\em et al.} measured 7 allowed transitions between the $2^3S$ 
and $2^3P$
manifolds.  These authors find  1.066$\pm$0.004 \cite{Pachucki12}, 
1.028$\pm$0.011 and 
1.074$\pm$0.003$fm^2$, respectively; the reason for the differences of  
several standard  deviations is  presently not understood.  
The shift from electron scattering agrees, but is  not
precise enough to favor one or the other of the values from atomic
Helium.  \\[5mm]
{\em Acknowledgement.~~} The author would like to thank Dirk Trautmann for helpful
discussions.
 
%
%

\begin{thebibliography}{10}

\bibitem{Antognini11}
A.~Antognini {\em et al.}
\newblock {\em Can. J. Physics}, 89:47, 2011.

\bibitem{Friar04}
J.L. Friar and I.~Sick.
\newblock {\em Phys. Lett. B}, 579:285, 2004.

\bibitem{Friar05b}
J.L. Friar and I.~Sick.
\newblock {\em Phys. Rev. A}, 72:040502, 2005.

\bibitem{Sick12}
I.~Sick.
\newblock {\em Prog. Part. Nucl. Phys.}, 67:473, 2012.

\bibitem{Pohl10a}
R.~Pohl, A.~Antognini, F.~Nez, F.D. Amaro, F.~Biraben, J.M.R. Cardoso, D.A.
  Covita, A.~Dax, S.~Dhawan, L.M.P. Fernandes, A.~Giesen, T.~Rraf, T.W.
  H{\"a}nsch, P.~Indelicato, L.~Julien, C-Y. Kao, P.~Knowles, J.A.M.Lopes,
  E-O.~Le Bigot, Y-W. Liu, L.~Ludhova, C.M.B. Monteiro, F.~Mulhauser, T.~Nebel,
  P.~Rabinowitz, J.M.F dos Santos, L.~Schaller, K.~Schuhmann, C.~Schwob,
  T.~Taqqu, J.F.C.A. Veloso, and F.~Kottmann.
\newblock {\em Nature}, 466:213, 2010.

\bibitem{Sick08}
I.~Sick.
\newblock {\em Phys. Rev. C}, 77:041302, 2008.

\bibitem{Frosch67}
R.~Frosch, J.S. McCarthy, R.E. Rand, and M.R. Yearian.
\newblock {\em Phys. Rev.}, 160:874, 1967.

\bibitem{Erich68}
U.~Erich, H.~Frank, D.~Haas, and H.~Prange.
\newblock {\em Z. Phys.}, 209:208, 1968.

\bibitem{McCarthy77}
J.S. McCarthy, I.~Sick, and R.R. Whitney.
\newblock {\em Phys. Rev. C}, 15:1396, 1977.

\bibitem{Arnold78}
R.G. Arnold, B.T. Chertok, S.~Rock, W.P. Schuetz, Z.M. Szalata, D.~Day, J.S.
  McCarthy, F.~Martin, B.A. Mecking, I.~Sick, and G.~Tamas.
\newblock {\em Phys. Rev. Lett.}, 40:1429, 1978.

\bibitem{vonGunten82}
A.~von {G}unten.
\newblock {\em Thesis, TH Darmstadt, unpublished}, 1982.

\bibitem{Ottermann85}
C.R. Ottermann, G.~Koebschall, K.~Maurer, K.~Roehrich, Ch. Schmitt, and V.H.
  Walther.
\newblock {\em Nucl. Phys. A}, 435:688, 1985.

\bibitem{Camsonne13}
A.~Camsonne {\em et al.}
\newblock {\em Phys. Rev. Lett.}, 112:132503, 2014.

\bibitem{Szalata77}
Z.M. Szalata, J.M. Finn, J.~Flanz, F.J. Kline, G.A.Peterson, J.W.~Lightbody
  Jr., X.K. Maruyama, and S.~Penner.
\newblock {\em Phys. Rev. C}, 15:1200, 1977.

\bibitem{Dunn83}
P.C. Dunn, S.B. Kowalski, F.N. Rad, C.P. Sargent, W.E. Turchinetz, R.~Goloskie,
  and D.P. Saylor.
\newblock {\em Phys. Rev. C}, 27:71, 1983.

\bibitem{McCarthy70}
J.S. McCarthy, I.~Sick, R.R. Whitney, and M.R. Yearian.
\newblock {\em Phys. Rev. Lett.}, 25:884, 1970.

\bibitem{Cavedon82}
J.M. Cavedon, B.~Frois, D.~Goutte, M.~Huet, Ph. Leconte, C.N. Papanicolas,
  X.-H. Phan, S.K. Platchkov, S.~Williamson, W.~Boeglin, and I.~Sick.
\newblock {\em Phys. Rev. Lett.}, 49:978, 1982.

\bibitem{Nakagawa01a}
I.~Nakagawa, J.~Shaw, S.~Churchwell, X.~Jiang, B.~Asavapibhop, M.C. Berisso,
  P.E. Bosted, K.~Burchesky, F.~Casagrande, A.~Cichocki, R.S. Hicks, A.~Hotta,
  T.~Kobayashi, R.A. Miskimen, G.A. Peterson, S.E. Rock, T.~Suda, T.~Tamae,
  W.~Turchinetz, and K.~Wang.
\newblock {\em Phys. Rev. Lett.}, 86:5446, 2001.

\bibitem{Collard65}
H.~Collard, R.~Hofstadter, E.B. Hughes, A.~Johanson, M.R. Yearian, R.B. Day,
  and R.T. Wagner.
\newblock {\em Phys. Rev. C}, 15:57, 1965.

\bibitem{Beck87a}
D.~Beck {\em et al.}
\newblock {\em Phys. Rev. Lett.}, 59:1537, 1987.

\bibitem{Sick14}
I.~Sick and D.~Trautmann.
\newblock {\em Phys. Rev. C}, 89:012201(R), 2014.

\bibitem{Brida11}
I.~Brida, S.C. Pieper, and R.B. Wiringa.
\newblock {\em Phys. Rev. C}, 84:024319, 2011.

\bibitem{Sauer00}
P.U. Sauer.
\newblock {\em priv. comm.}, 2000.

\bibitem{VanRooij11}
R.~van Rooij, J.S. Borbely, J.~Simoneet, M.D. Hoogerland, K.S.E. Eikerma, R.A.
  Rozendaal, and W.~Wassen.
\newblock {\em Science}, 333:196, 2011.

\bibitem{Pastor12}
P.~{Cancio Pastor}, L.~Consolino, G.~Giusfredi, P.~De Natale, M.~Inguscio, V.A.
  Yerekhin, and K.~Pachucki.
\newblock {\em Phys. Rev. Lett.}, 108:143001, 2012.

\bibitem{Pachucki12}
K.~Pachucki, V.A. Yerokhin, and P.~Cancio Pastor.
\newblock {\em Phys. Rev. A}, 85:042517, 2012.

\bibitem{Shiner95}
D.~Shiner, R.~Dixson, and V.~Vedantham.
\newblock {\em Phys. Rev. Lett.}, 74:3553, 1995.

\end{thebibliography}

\end{document}